\documentclass{article}
\usepackage[T1]{fontenc}

\makeatletter

\newcommand{\LyX}{L\kern-.1667em\lower.25em\hbox{Y}\kern-.125emX\@}

\makeatother

\begin{document}

\title{Quantum magnets: a brief overview }

\author{Indrani Bose }

\maketitle
\begin{center}

Department of Physics

Bose Institute

93/1, A. P. C. Road

Calcutta-700009

\end{center}

\begin{abstract}
Quantum magnetism is one of the most active areas of research in condensed matter
physics. There is significant research interest specially in low-dimensional
quantum spin systems. Such systems have a large number of experimental realizations
and exhibit a variety of phenomena the origin of which can be attributed to
quantum effects and low dimensions. In this review, an overview of some aspects
of quantum magnetism in low dimensions is given. The emphasis is on key concepts,
theorems and rigorous results as well as models of spin chains, ladders and
frustated magnetic systems.
\end{abstract}

\section{Introduction}

Quantum magnets are spin systems in which the spins interact via the well-known
exchange interaction. The interaction is purely quantum mechanical in nature
and the form of the interaction was derived simultaneously by Heisenberg and
Dirac in 1926 \cite{1}. The most well-known model of interacting spins in an
insulating solid is the Heisenberg model with the Hamiltonian

\begin{equation}
\label{1}
H=\sum _{\left\langle ij\right\rangle }J_{ij}\overrightarrow{S_{i}}.\overrightarrow{S_{j}}
\end{equation}
 \( \overrightarrow{S_{i}} \) is the spin operator located at the lattice site
\( i \) and \( J_{ij} \) denotes the strength of the exchange interaction.
The spin \( \left| \overrightarrow{S_{i}}\right|  \) can have a magnitude 1/2,
1, 3/2, 2, ...etc. The lattice, at the sites of which the spins are located,
is d-dimensional. Examples are a linear chain (d = 1), the square lattice (d
= 2 ) and the cubic lattice (d =3 ). Ladders have structures interpolating between
the chain and the square lattice. An n-chain ladder consists of n chains (n
= 2, 3, 4,...etc.) coupled by rungs. Real magnetic solids are three-dimensional
(3d) but can be effectively considered as low-dimensional systems if the exchange
interactions have different strengths in different directions. To give an example,
a magnetic solid may consist of chains of spins. The solid may be considered
as a linear chain (d=1) compound if the intra-chain exchange interactions are
much stronger than the inter-chain ones. In a planar (d=2) magnetic system,
the dominant exchange interactions are intra-planar. Several examples of low-dimensional
magnetic systems are given in \cite{2}. 

The strength of the exchange interaction \( J_{ij} \) in Eq.(1) falls down
rapidly as the distance between interacting spins increases. For many solids,
the sites i and j are nearest-neighbours (n.n.s) on the lattice and \( J_{ij} \)
's have the same magnitude J for all the n.n. interactions. The Hamiltonian
in (1) then becomes 

\begin{equation}
\label{2}
H=J\sum _{\left\langle ij\right\rangle }\overrightarrow{S_{i}}.\overrightarrow{S_{j}}
\end{equation}
 There are, however, examples of magnetic systems in which the strengths of
the exchange interactions between successive pairs of spins are not the same.
Also, the interaction Hamiltonian (1) may include n.n. as well as further-neighbour
interactions. The well-known Majumdar-Ghosh chain \cite{3} is described by
the Hamiltonian

\begin{equation}
\label{3}
H_{MG}=J\sum ^{N}_{i=1}\overrightarrow{S_{i}}.\overrightarrow{S}_{i+1}+\frac{J}{2}\sum ^{N}_{i=1}\overrightarrow{S_{i}}.\overrightarrow{S}_{i+2}
\end{equation}
 and includes n.n. as well as next-nearest-neighbour (n.n.n.) interactions.The
Haldane-Shastry model \cite{4} has a Hamiltonian of the form

\begin{equation}
\label{4}
H=J\sum _{\left\langle ij\right\rangle }\frac{1}{\left| i-j\right| ^{2}}\overrightarrow{S_{i}}.\overrightarrow{S_{j}}
\end{equation}
 and includes long-ranged interactions. Real materials are characterised by
various types of anisotropy. The fully anisotropic n.n. Heisenberg Hamiltonian
in 1d is given by

\begin{equation}
\label{5}
H_{XYZ}=\sum ^{N}_{i=1}[J_{x}S^{x}_{i}S^{x}_{i+1}+J_{y}S^{y}_{i}S^{y}_{i+1}+J_{z}S^{z}_{i}S^{z}_{i+1}]
\end{equation}
 The special cases of this Hamiltonian are: the Ising (\( J_{x}=J_{y}=0 \))
, the \( XY \) (\( J_{z}=0 \)), the XXX or isotropic Heisenberg (\( J_{x}=J_{y}=J_{z} \)
) and the XXZ or anisotropic Heisenberg (\( J_{x}=J_{y}\neq J_{z} \) ) models.
There is a huge literature on these models some of which are summarised in Refs.
\cite{1,2,5,6,7}. Other anisotropy terms may be present in the full spin Hamiltonian
besides the basic exchange interaction terms. 

Consider the isotropic Heisenberg Hamiltonian in (2) where \( \left\langle ij\right\rangle  \)
denotes a n.n. pair of spins. The sign of the exchange interaction determines
the favourable alignment of the n.n. spins. \( J>0(J<0) \) corresponds to antiferromagnetic
(ferromagnetic) exchange interaction. To see how exchange interaction leads
to magnetic order, treat the spins as classical vectors. Each n.n. spin pair
has an interaction energy \( JS^{2}cos\theta  \) where \( \theta  \) is the
angle between n.n. spin orientations. When J is < 0, the lowest energy is achieved
when \( \theta =0 \), i.e., the interacting spins are parallel. The ferromagnetic
(FM) ground state has all the spins parallel and the ground state energy \( E_{g}=-J\frac{NzS^{2}}{2} \)
where z is the coordination number of the lattice. When \( J \) is > 0, the
lowest energy is achieved for \( \theta =\pi  \) , i.e., the n.n. spins are
antiparallel. The antiferromagnetic (AFM) ground state is the N\'{e}el state
in which n.n. spins are antiparallel to each other. The ground state energy
\( E_{g}=-\frac{JNzS^{2}}{2} \). 

Magnetism , however, is a purely quantum phenomenon and the Hamiltonian (2)
is to be treated quantum mechanically rather than classically. For simplicity,
consider a chain of spins of magnitude \( \frac{1}{2} \). Periodic boundary
condition is assumed, i.e., \( \overrightarrow{S}_{N+1}=\overrightarrow{S_{1}} \)
. The Hamiltonian (2) can be written as

\begin{equation}
\label{6}
H=J\sum ^{N}_{i=1}[S^{z}_{i}S^{z}_{i+1}+\frac{1}{2}(S^{+}_{i}S^{-}_{i+1}+S^{-}_{i}S^{+}_{i+1})]
\end{equation}
where

\begin{equation}
\label{7}
S^{\pm }_{i}=S^{x}_{i}\pm iS^{y}_{i}
\end{equation}
are the raising and lowering operators. It is easy to check that in the case
of a FM, the classical ground state is still the quantum mechanical ground state
with the same ground state energy. However, the classical AFM ground state (the
N\'{e}el state) is not the quantum mechanical ground state. The determination
of the exact AFM ground state is a tough many body problem and the solution
can be obtained with the help of the Bethe Ansatz technique (Section 2).

For a spin-1/2 system, the number of eigenstates of the Hamiltonian is \( 2^{N} \)
where N is the number of spins. In a real solid N is \( \sim 10^{23} \) and
exact determination of all the eigenvalues and the eigenfunctions of the system
is impossible. There are some classes of AFM spin models for which the ground
state and in a few cases the low-lying excitation spectrum are known exactly
(Section 2). In the majority of cases, however, the ground state and the low-lying
excited states are determined in an approximate manner. Knowledge of the low-lying
excitation spectrum enables one to determine the low-temperature thermodynamics
and the response to weak external fields. The usual thermodynamic quantities
of a magnetic system are magnetisation, specific heat and susceptibility. Exchange
interaction can give rise to magnetic order below a critical temperature. However,
for some spin systems, the ground state is disordered, i.e., there is no magnetic
order even at T = 0. Long range order (LRO) of the N\'{e}el-type exists in the
magnetic system if

\begin{equation}
\label{8}
lim_{R\rightarrow \infty }\left\langle \overrightarrow{S}(0).\overrightarrow{S}(\overrightarrow{R})\right\rangle \neq 0
\end{equation}
 where \( \overrightarrow{R} \) denotes the spatial location of the spin. At
\( T=0 \), the expectation value is in the ground state and at \( T\neq 0 \)
, the expectation value is the usual thermodynamic average. The dynamical properties
of a magnetic system are governed by the time-dependent pair correlation functions
or their Fourier transforms. Quantities of experimental interest include the
dynamical correlation functions in neutron scattering experiments, the NMR spin-lattice
relaxation rate, various relaxation functions and associated lineshapes as well
as the dynamical response of the magnetic system to various spectroscopic probes
\cite{8}. Knowledge of the ground and low-lying states and the corresponding
energy eigenvalues is essential to determine the thermodynamic and dynamic properties
of a magnetic system.

The discovery of high-temperature cuprate superconductors in 1987 has given
a tremendous boost to research activity in magnetism. The dominant electronic
and magnetic properties of the cuprate systems are associated with the copper-oxide
(\( CuO_{2}) \) planes. The \( Cu^{2+} \) ions carry spin-\( \frac{1}{2} \)
and the spins interact via the Heisenberg AFM exchange interaction Hamiltonian.
This fact has given rise to a large number of studies on 2d antiferromagnets.
The cuprates exhibit a variety of novel phenomena in their insulating, metallic
and superconducting phases some of which at least have links to quantum magnetism.
The subject of magnetism has, as a result, expanded significantly in scope and
content. A rich interplay between theory and experiments has led to the discovery
of materials exhibiting hitherto unknown phenomena, formulation of new theoretical
ideas, solution of old puzzles and opening up of new research possibilities.
In this review, a brief overview of some of the important developments in quantum
magnetism will be given. The focus is on quantum antiferromagnets and insulating
solids.

\section{Theorems and rigorous results}

(i) Theorems :\\
A. Lieb-Mattis theorem \cite{9}

For general spin and for all dimensions and also for a bipartite lattice, the
entire eigenvalue spectrum satisfies the inequality

\begin{equation}
\label{9}
E_{0}(S)\leq E_{0}(S+1)
\end{equation}
 where \( E_{0}(S) \) is the minimum energy corresponding to total spin S.
The weak inequality becomes a strict inequality for a FM exchange coupling between
spins of the same sublattice. The theorem is valid for any range of exchange
coupling and the proof does not require PBC. The ground state of the \( S=\frac{1}{2} \)
Heisenberg AFM with an even number N of spins is a singlet according to the
Lieb-Mattis theorem.\\
B. Marshall's sign rule \cite{10}

The rule specifies the structure of the ground state of a n.n. \( S=\frac{1}{2} \)
Heisenberg Hamiltonian defined on a bipartite lattice. The rule can be generalised
to spin S, n.n.n. FM interaction but not to n.n.n. AFM interaction. A bipartite
lattice is a lattice which can be divided into two sublattices A and B such
that the n.n. spins of a spin belonging to the A sublattice are located in the
B sublattice and vice versa. Examples of such lattices are the linear chain,
the square and the cubic lattices. According to the sign rule, the ground state
\( \psi  \) has the form

\begin{equation}
\label{10}
\left| \psi \right\rangle =\sum _{\mu }C_{\mu }\left| \mu \right\rangle 
\end{equation}
 where \( \left| \mu \right\rangle  \) is an Ising basis state. The coefficient
\( C_{\mu } \) has the form

\begin{equation}
\label{11}
C_{\mu }=(-)^{p_{\mu }}a_{\mu }
\end{equation}
 with \( a_{\mu } \) real and \( \geq 0 \) and \( p_{\mu } \) is the number
of up-spins in the A sublattice.\\
C. Lieb, Schultz and Mattis (LSM) theorem \cite{11}:

A half-integer S spin chain described by a reasonably local Hamiltonian respecting
translational and rotational symmetry either has gapless excitation spectrum
or has degenerate ground states, corresponding to spontaneously broken translational
symmetry.\\
In the case of a gapless excitation spectrum, there is at least one momentum
wave vector for which the excitation energy is zero. For a spectrum with gap,
the lowest excitation is separated from the ground state by an energy gap \( \Delta  \).
The temperature dependence of thermodynamic quantities is determined by the
nature of the excitation spectrum (with or without gap). The LSM theorem does
not hold true for integer spin chains. For such chains, Haldane made a conjecture
that the spin excitation spectrum is gapped \cite{12}. This conjecture has
been verified both theoretically and experimentally \cite{13}.\\
D. Oshikawa, Yamanaka and Affleck theorem \cite{14}

This theorem extends the LSM theorem to the case of an applied magnetic field.
The content of the theorem is : translationally invariant spin chains in an
applied field can have a gapped excitation spectrum, without breaking translational
symmetry, only when the magnetization per site m (\( m=\frac{1}{N}\sum ^{N}_{i=1}S^{z}_{i} \),
\( N \) is the total number of spins in the system ) obeys the relation

\begin{equation}
\label{12}
S-m=integer
\end{equation}
 where \( S \) is the magnitude of the spin. The proof is an easy extension
of that of the LSM theorem. The gapped phases correspond to magnetization plateaux
in the m vs. H curve at the quantized values of m which satisfy (12). Whenever
there is a gap in the spin excitation spectrum, it is obvious that the magnetization
cannot change in changing external field. Fractional quantization can also occur,
if accompanied by (explicit or spontaneous) breaking of the translational symmetry.
In this case, the plateau condition is given by

\begin{equation}
\label{13}
n(S-m)=integer
\end{equation}
 where n is the period of the ground state. Hida \cite{15} has considered a
\( S=\frac{1}{2} \) HAFM chain with period 3 exchange coupling. A plateau in
the magnetization curve occurs at \( m=\frac{1}{6} \) (\( \frac{1}{3} \) of
full magnetization ). In this case, n =3, \( S=\frac{1}{2} \) and \( m=\frac{1}{6} \)
and the quantization condition in (13) is obeyed. Ref. \cite{16} gives a review
of magnetization plateaux in interacting spin systems. Magnetization plateaux
have been observed in the magnetic compound \( NH_{4}CuCl_{3} \) at \( m=\frac{1}{4} \)
and \( \frac{3}{4} \) \cite{17}. Possible extensions of the LSM theorem to
higher dimensions have been suggested \cite{18}. The compound \( SrCu_{2}(BO_{3})_{2} \)
is the first AFM compound in 2d in which magnetization plateaux have been observed
experimentally \cite{19}. Like the quantum Hall effect, the phenomenon of magnetization
plateaux is another striking example of the quantization of a physically measurable
quantity as a function of the magnetic field.\\
E. Mermin-Wagner's theorem \cite{20}  

There cannot be any AFM LRO at finite T in dimensions d =1 and 2. The LRO can,
however, exist in the ground state of spin models in d =2. LRO exists in the
ground state of the 3d HAFM model for spin \( S\geq \frac{1}{2} \) \cite{21}.
At finite T, the LRO persists upto a critical temperature \( T_{c} \) . For
square \cite{22} and hexagonal \cite{23} lattices, LRO exists in the ground
state for \( S\geq 1 \) . The above results are based on rigorous proofs. No
such proof exists as yet for \( S=\frac{1}{2} \) , d =2 (this case is of interest
because the \( CuO_{2} \) plane of the high-\( T_{c} \) cuprate systems is
a \( S=\frac{1}{2} \) 2d AFM).\\
(ii) Exact Results :\\
A. the Bethe Ansatz \cite{24}

The Bethe Ansatz (BA) was formulated by Bethe in 1931 and describes a wave function
with a particular kind of structure. Bethe considered the spin\( -\frac{1}{2} \)
Heisenberg linear chain in which only n.n. spins interact. In the case of the
FM chain, the exact ground state is simple with all spins aligned in the same
direction, say, pointing up. An excitation is created in the system by deviating
a spin from its ground state arrangement, i.e., replacing an up-spin by a down-spin.
Due to the exchange interaction, the deviated spin does not stay localised at
a particular site but travels along the chain of spins. This excitation is the
so-called spin wave or magnon. For the isotropic FM Heisenberg Hamiltonian,
the exact one-magnon eigenstate is given by

\begin{equation}
\label{14}
\psi =\sum ^{N}_{m=1}e^{ikm}S^{-}_{m}\left| \uparrow \uparrow \uparrow .......\right\rangle 
\end{equation}
 where m denotes the site at which the down-spin is located and the summation
over m runs from the first to the last site in the chain. The k's are the ``momenta''
which from periodic boundary conditions have N allowed values

\begin{equation}
\label{15}
k=\frac{2\pi }{N}\lambda ,\lambda =0,1,2,...,N-1
\end{equation}
 The excitation energy \( \epsilon _{k} \), measured w.r.t. the ground state
energy and in units of J, is

\begin{equation}
\label{16}
\epsilon _{k}=(1-cosk)
\end{equation}
 In the case of r spin deviations (magnons), the eigenfunction can be written
as

\begin{equation}
\label{17}
\psi ^{(r)}=\sum _{m_{1}<m_{2}<...m_{r}}a(m_{1,}m_{2},...,m_{r})S^{-}_{m_{1}}S^{-}_{m_{2}}....S^{-}_{m_{r}}\left| \uparrow \uparrow \uparrow .......\right\rangle 
\end{equation}
 The amplitudes are given by the BA

\begin{equation}
\label{18}
a(m_{1},m_{2},...,m_{r})=\sum _{P}e^{i\sum _{j}k_{Pj}m_{j}+\frac{1}{2}i\sum ^{1,r}_{j<l}\phi _{Pj,Pl}}
\end{equation}
 where \( P \) stands for a permutation of the set \{1,2,...,r\} and \( Pj \)
is the image of \( j \) under permutation. The sum is over all the \( r! \)
permutations. Each term in (18) describes r plane waves. The scattering of a
pair of waves introduces the phase shift \( \phi _{jl} \) . The symmetric sum
over permutations is in accordance with the bosonic nature of the waves, the
spin waves, propagating along the chain. The energy of the state \( \psi ^{(r)} \)
is

\begin{equation}
\label{19}
\epsilon ^{(r)}=\sum ^{r}_{i=1}(1-cosk_{i})
\end{equation}
 The k's are determined as before by applying the periodic boundary conditions
which leads to the r equations

\begin{equation}
\label{20}
Nk_{i}=2\pi \lambda _{i}+\sum _{j}\phi _{ij}
\end{equation}
where \( \lambda _{i} \) 's are r integers. One further imposes the condition
that a spin at a particular site cannot be deviated more than once leading to
the relations

\begin{equation}
\label{21}
2cot\frac{1}{2}\phi _{ij}=cot\frac{k_{i}}{2}-cot\frac{k_{j}}{2}
\end{equation}
 Since \( \phi _{ij}=-\phi _{ji} \) , Eqs. (21) are \( \frac{r(r-1)}{2} \)
in number, i.e., there are as many distinct \( \phi  \) 's. Eqs. (20) are r
in number. Together, the total number is \( \frac{r(r+1)}{2} \) equations in
as many unknowns. Bethe thus established that the set of equations could be
expected to have solutions. 

The momenta \( k_{i} \) 's can be real or complex. In the first case, the spin
waves or magnons scatter against each other giving rise to a continuum of scattering
states. In the second case, the magnons form bound states, i.e., the reversed
spins tend to be located at n.n. lattice positions. For r magnons, the r-magnon
bound state energy is given by

\begin{equation}
\label{22}
\epsilon =\frac{1}{r}(1-cosK)
\end{equation}
 where \( K=\sum ^{r}_{i=1}k_{i} \) is the total centre of mass momentum of
r magnons. The results can be generalised to the XXZ Heisenberg Hamiltonian.
The multimagnon bound states were first detected in the quasi-one-dimensional
magnetic system \( CoCl_{2}.2H_{2}O \) at pumped helium temperatures and in
high magnetic fields by far infrared spectroscopy \cite{25}. Later improvements
\cite{26} made use of infrared HCN/DCN lasers, the high intensity of which
made possible observation of even 14 magnon bound states.

The exact ground state energy of the isotropic Heisenberg Hamiltonian (Eq.(2))
can be determined using the BA. The BA equations are the same as in the FM case
but the sign of the exchange integral changes from \( -J \) to \( J \) (\( J>0 \)
). The total spin of the AFM ground state is \( S=0 \) according to the Lieb-Mattis
theorem. In the ground state, \( \frac{N}{2} \) spins are up and \( \frac{N}{2} \)
spins down \( (r=\frac{N}{2}) \). The ground stae is non-degenerate and there
is a unique choice of the \( \lambda _{i} \) 's as

\begin{equation}
\label{23}
\lambda _{1}=1,\lambda _{2}=3,\lambda _{3}=5,....,\lambda _{\frac{N}{2}}=N-1
\end{equation}
The ground state is also spin disordered, i.e., has no AFM LRO. The exact ground
state energy \( E_{g} \) is

\begin{equation}
\label{24}
E_{g}=\frac{NJ}{4}-JNln2
\end{equation}
 The low-lying excitation spectrum has been calculated by des Cloizeaux and
Pearson (dCP) \cite{27} by making appropriate changes in the distribution of
\( \lambda _{i} \) 's in the ground state. The spectrum is given by

\begin{equation}
\label{25}
\epsilon =\frac{\pi }{2}\left| sink\right| ,-\pi \leq k\leq \pi 
\end{equation}
for spin 1 states. The wave vector \( k \) is measured w.r.t. that of the ground
state. A more rigorous calculation of the low-lying excitation spectrum has
been given by Faddeev and Takhtajan \cite{28}. There are \( S=1 \) as well
as \( S=0 \) states. We give a qualitative description of the excitation spectrum,
for details Ref. \cite{28} should be consulted. The energy of the low-lying
excited states can be written as \( E(k_{1},k_{2})=\epsilon (k_{1})+\epsilon (k_{2}) \)
with \( \epsilon (k_{i})=\frac{\pi }{2}sink_{i} \) and total momentum \( k=k_{1}+k_{2} \).
At a fixed total momentum \( k \), one gets a continuum of scattering states.
The lower boundary of the continuum is given by the dCP spectrum (one of the
\( k_{i}'s=0 \) ). The upper boundary is obtained for \( k_{1}=k_{2}=\frac{k}{2} \)
and

\begin{equation}
\label{26}
\epsilon ^{U}_{k}=\pi \left| sin\frac{k}{2}\right| 
\end{equation}
 The energy-momentum relations suggest that the low-lying spectrum is actually
a combination of two elementary excitations known as spinons. The energies and
the momenta of the spinons just add up, showing that they do not interact. A
spinon is a \( S=\frac{1}{2} \) object, so on combination they give rise to
both \( S=1 \) and \( S=0 \) states. In the Heisenberg model, the spinons
are only noninteracting in the thermodynamic limit \( N\rightarrow \propto  \).
For an even number \( N \) of sites, the total spin is always an integer, so
that the spins are always excited in pairs. The spinons can be visualised as
kinks in the AFM order parameter. Due to the exchange interaction, the individual
spinons get delocalised into plane wave states. Inelastic neutron scattering
study of the linear chain \( S=\frac{1}{2} \) HAFM compound \( KCuF_{3} \)
has confirmed the existence of unbound spinon pair excitations \cite{29}. The
Haldane-Shastry model \cite{4} is another spin\( -\frac{1}{2} \) model in
1d for which the ground state and low-lying excitation spectrum are known exactly.
The ground state has the same functional form as the fractional quantum Hall
ground state and is spin-disordered. The elementary excitations are spinons
which are noninteracting even away from the thermodynamic limit, i.e., in finite
systems. The individual spinons behave as semions, i.e., have statistical properties
intermediate between fermions and bosons. In the case of integer spin chains,
the spinons are bound and the excitation spectrum consists of spin-wave-like
modes exhibiting the Haldane gap. The BA technique described in this Section
is the one originally proposed by Bethe. There is an algebraic version of the
BA which is in wide use and which gives the same final results as the earlier
technique. For an introduction to the algebraic BA method, see the Refs. \cite{30,31}.
A tutorial review of the BA is given in Ref. \cite{32}. The BA was originally
proposed for the Heisenberg model in magnetism. Later, the method was applied
to other interacting many body systems in 1d such as the Fermi and Bose gas
models in which particles on a line interact through delta function potentials
\cite{33}, the Hubbard model in 1d \cite{34}, 1d plasma which crystallizes
as a Wigner solid \cite{35}, the Lai-Sutherland model \cite{36} which includes
the Hubbard model and a dilute magnetic model as special cases, the Kondo model
in 1d \cite{37}, the single impurity Anderson model in 1d \cite{38}, the supersymmetric
t-J model (J = 2t) \cite{39} etc. In the case of quantum models, the BA method
is applicable only to 1d models. The BA method has also been applied to derive
exact results for classical lattice statistical models in 2d.\\
B. The Majumdar-Ghosh chain \cite{3,7}\\
The Hamiltonian is given in Eq. (3). The exact ground state of \( H_{MG} \)
is doubly degenerate and the states are

\begin{equation}
\label{27}
\phi _{1}\equiv [12][34]...[N-1N],\phi _{2}\equiv [23][45]...[N1]
\end{equation}
 where \( \left[ lm\right]  \) denotes a singlet spin configuration for spins
located at the sites \( l \) and \( m \). Also, PBC is assumed. One finds
that translational symmetry is broken in the ground state. The proof that \( \phi _{1} \)
and \( \phi _{2} \) are the exact ground states can be obtained by the method
of `divide and conquer'. One can verify that \( \phi _{1} \) and \( \phi _{2} \)
are exact eigenstates of \( H_{MG} \) by applying the spin identity \( \overrightarrow{S}_{n}.(\overrightarrow{S}_{l}+\overrightarrow{S}_{m})[lm]=0 \)
. Let \( E_{1} \) be the energy of \( \phi _{1} \) and \( \phi _{2} \) .
Let \( E_{g} \) be the exact ground state energy. Then \( E_{g}\leq E_{1} \)
. One divides the Hamiltonian \( H \) into sub-Hamiltonians , \( H_{i} \)
's, such that \( H=\sum _{i}H_{i} \) . \( H_{i} \) can be exactly diagonalised
and let \( E_{i0} \) be the ground state energy. Let \( \psi _{g} \) be the
exact ground state wave function. By variational theory,

\begin{equation}
\label{28}
E_{g}=\left\langle \psi _{g}\left| H\right| \psi _{g}\right\rangle =\sum _{i}\left\langle \psi _{g}\left| H_{i}\right| \psi _{g}\right\rangle \geq \sum _{i}E_{i0}
\end{equation}
 One thus gets,

\begin{equation}
\label{29}
\sum _{i}E_{i0}\leq E_{g}\leq E_{1}
\end{equation}
 If one can show that \( \sum _{i}E_{i0}=E_{1} \) , then \( E_{1} \) is the
exact ground state energy. For the MG-chain, the sub-Hamiltonian \( H_{i} \)
is 

\begin{equation}
\label{30}
H_{i}=\frac{J}{2}(\overrightarrow{S}_{i}.\overrightarrow{S}_{i+1}+\overrightarrow{S}_{i+1}.\overrightarrow{S}_{i+2}+\overrightarrow{S}_{i+2}.\overrightarrow{S}_{i})
\end{equation}
 There are \( N \) such sub-Hamiltonians. One can easily verify that \( E_{i0}=-\frac{3J}{8} \)
and \( E_{1}=-\frac{3J}{4}\frac{N}{2} \) ( -\( \frac{3J}{4} \) is the energy
of a singlet and there are \( \frac{N}{2} \) VBs in \( \phi _{1} \) and \( \phi _{2} \)).
From (29), one finds that the lower and upper bounds of \( E_{g} \) are equal
and hence \( \phi _{1} \) and \( \phi _{2} \) are the exact ground states
with energy \( E_{1}=-\frac{3JN}{8} \) . There is no LRO in the two-spin correlation
function in the ground stae:

\begin{equation}
\label{31}
K^{2}(i,j)=\left\langle S^{z}_{i}S^{z}_{j}\right\rangle =\frac{1}{4}\delta _{ij}-\frac{1}{8}\delta _{\left| i-j\right| ,1}
\end{equation}
 The four-spin correlation function has off-diagonal LRO.

\begin{eqnarray}
K^{4}(ij,lm) & = & \left\langle S^{x}_{i}S^{x}_{j}S^{y}_{l}S^{y}_{m}\right\rangle \nonumber \\
 & = & K^{2}(ij)K^{2}(lm)\nonumber \\
 & + & \frac{1}{64}\delta _{\left| i-j\right| ,1}\delta _{\left| l-m\right| ,1}exp(i\pi (\frac{i+j}{2}-\frac{l+m}{2}))\label{32} 
\end{eqnarray}
 Let \( T \) be the translation operator for unit displacement. Then

\begin{equation}
\label{33}
T\phi _{1}=\phi _{2},T\phi _{2}=\phi _{1}
\end{equation}
 The states 

\begin{equation}
\label{34}
\phi ^{+}=\frac{1}{\sqrt{2}}(\phi _{1}+\phi _{2}),\phi ^{-}=\frac{1}{\sqrt{2}}(\phi _{1}-\phi _{2})
\end{equation}
 correspond to momentum wave vectors \( k=0 \) and \( k=\pi  \) . The excitation
spectrum is not exactly known. Shastry and Sutherland \cite{40} have derived
the excitation spectrum in the basis of `defect' states. A defect state has
the wave function

\begin{equation}
\label{35}
\psi (p,m)=...[2p-3,2p-2]\alpha _{2p-1}[2p,2p+1]...[2m-2,2m-1]\alpha _{2m}[2m+1,2m+2]...
\end{equation}
 where the defects (\( \alpha _{2p-1} \) and \( \alpha _{2m} \)) separate
two ground states. The two defects are up-spins and the total spin of the state
is 1. Similarly, the defect spins can be in a singlet spin configuration so
that the total spin of the state is 0. Because of PBC, the defects occur in
pairs. A variational state can be constructed by taking a linear combination
of the defect states. The excitation spectrum consists of a continuum with a
lower edge at \( J(\frac{5}{2}-2\left| cosk\right| ) \). A bound state of the
two defects can occur in a restricted region of momentum wave vectors. The MG
chain has been studied for general values \( \alpha J \) of the n.n.n. interaction
\cite{41}. The ground state is known exactly only at the MG point \( \alpha =\frac{1}{2} \)
. The excitation spectrum is gapless for \( 0<\alpha <\alpha _{cr}(\simeq 0.2411) \).
Generalizations of the MG model to two dimensions exist \cite{42,43}. The Shastry-Sutherland
model \cite{42} is defined on a square lattice and includes diagonal interactions
as shown in Figure 1. The n.n. and diagonal exchange interactions are of strength
\( J_{1} \) and \( J_{2} \) respectively. For \( \frac{J_{1}}{J_{2}} \) below
a critical value \( \sim 0.7 \) , the exact ground state consists of singlets
along the diagonals. At the critical point, the ground state changes from the
gapful disordered state to the AFM ordered gapless state. The compound \( SrCu_{2}(BO_{3})_{2} \)
is well-described by the Shastry-Sutherland model \cite{19}. Bose and Mitra
\cite{43} have constructed a \( J_{1}-J_{2}-J_{3}-J_{4}-J_{5} \) spin-\( \frac{1}{2} \)
model on the square lattice. \( J_{1},J_{2},J_{3},J_{4} \) and \( J_{5} \)
are the strengths of the n.n., diagonal, n.n.n., knight's-move-distance-away
and further-neighbour-diagonal exchange interactions (Figure 2). The four columnar
dimer states (Figure 3) have been found to be the exact eigenstates of the spin
Hamiltonian for the ratio of interaction strengths

\begin{equation}
\label{36}
J_{1}:J_{2}:J_{3}:J_{4}:J_{5}=1:1:\frac{1}{2}:\frac{1}{2}:\frac{1}{4}
\end{equation}
It has not been possible as yet to prove that the four columnar dimer states
are also the ground states. Using the method of `divide and conquer', one can
only prove that a single dimer state is the exact ground state with the dimer
bonds of strength \( 7J \). The strengths of the other exchange interactions
are as specified in (36). For a \( 4\times 4 \) lattice with PBC, one can trivially
show that the four CD states are the exact ground states. 

C. The Affleck-Kennedy-Lieb-Tasaki (AKLT) model \cite{44}

We have already discussed the LSM theorem, the proof of which fails for integer
spin chains. Haldane \cite{12,13} in 1983 made the conjecture, based on a mapping
of the HAFM Hamiltonian, in the long wavelength limit, onto the nonlinear \( \sigma  \)
model, that integer-spin HAFM chains have a gap in the excitation spectrum.
The conjecture has now been verified both theoretically and experimentally \cite{45}.
In 1987, AKLT constructed a spin-1 model in 1d for which the ground state could
be determined exactly \cite{44}. Consider a 1d lattice, each site of which
is occupied by a spin-1. Each such spin can be considered to be a symmetric
combination of two spin-\( \frac{1}{2} \) 's. Thus, one can write down

\begin{eqnarray}
\psi _{++} & = & \left| ++\right\rangle ,S^{z}=+1\nonumber \\
\psi _{--} & = & \left| --\right\rangle ,S^{z}=-1\nonumber \\
\psi _{+-} & = & \frac{1}{\sqrt{2}}(\left| +-\right\rangle +\left| -+\right\rangle ,S^{z}=0\nonumber \\
\psi _{-+} & = & \psi _{+-}\label{37} 
\end{eqnarray}
 where `+' (`\( - \)') denotes an up (down) spin.

AKLT constructed a valence bond solid (VBS) state in the following manner. In
this state, each spin-\( \frac{1}{2} \) component of a spin-1 forms a singlet
(valence bond) with a spin-\( \frac{1}{2} \) at a neighbouring site. Let \( \epsilon ^{\alpha \beta } \)
(\( \alpha ,\beta =+ \) or \( - \)) be the antisymmetric tensor:

\begin{equation}
\label{38}
\epsilon ^{++}=\epsilon ^{--}=0,\epsilon ^{+-}=-\epsilon ^{-+}=1
\end{equation}
 A singlet spin configuration can be expressed as \( \frac{1}{\sqrt{2}}\epsilon ^{\alpha \beta }\left| \alpha \beta \right\rangle  \)
, summation over repeated indices being implied. The VBS wave function (with
PBC) can be written as 

\begin{equation}
\label{39}
\left| \psi _{VBS}\right\rangle =2^{-\frac{N}{2}}\psi _{\alpha _{1}\beta _{1}}\epsilon ^{\beta _{1}\alpha _{2}}\psi _{\alpha _{2}\beta _{2}}\epsilon ^{\beta _{2}\alpha _{3}}.....\psi _{\alpha _{i}\beta _{i}}\epsilon ^{\beta _{i}\alpha _{i+1}}\psi _{\alpha _{N}\beta _{N}}\epsilon ^{\beta _{N}\alpha _{1}}
\end{equation}
 \( \left| \psi _{VBS}\right\rangle  \) is a linear superposition of all configurations
in which each \( S^{z}=+1 \) is followed by a \( S^{z}=-1 \) with an arbitrary
number of \( S^{z}=0 \) spins in between and vice versa. If one leaves out
the zero's, one gets a N\'{e}el-type of order. One can define a non-local string
operator

\begin{equation}
\label{40}
\sigma ^{\alpha }_{ij}=-S^{\alpha }_{i}exp(i\pi \sum ^{j-1}_{l=i+1}S^{\alpha }_{l})S^{\alpha }_{j},(\alpha =x,y,z)
\end{equation}
and the order parameter 

\begin{equation}
\label{41}
O^{\alpha }_{string}=lim_{\left| i-j\right| \rightarrow \infty }\left\langle \sigma ^{\alpha }_{ij}\right\rangle 
\end{equation}
The VBS state has no conventional LRO but is characterised by a non-zero value
\( \frac{4}{9} \) of \( O^{\alpha }_{string} \). After constructing the VBS
state, AKLT determined the Hamiltonian for which the VBS state is the exact
ground state. The Hamiltonian is

\begin{equation}
\label{42}
H_{AKLT}=\sum _{i}P_{2}(\overrightarrow{S}_{i}+\overrightarrow{S}_{i+1})
\end{equation}
 where \( P_{2} \) is the projection operator onto spin 2 for a pair of n.n.
spins. The presence of a VB between each neighbouring pair implies that the
total spin of each pair cannot be 2 (after two of the \( S=\frac{1}{2} \) variables
form a singlet, the remaining \( S=\frac{1}{2} \) 's could form either a triplet
or a singlet). Thus, \( H_{AKLT} \) acting on \( \left| \psi _{VBS}\right\rangle  \)
gives zero. Since \( H_{AKLT} \) is a sum over projection operators, the lowest
possible eigenvalue is zero. Hence, \( \left| \psi _{VBS}\right\rangle  \)
is the ground state of \( H_{AKLT} \) with eigenvalue zero. The AKLT ground
state (the VBS state) is spin-disordered and the two-spin correlation function
has an exponential decay. The total spin of two spin-1's is 2, 1, 0. The projection
operator onto spin \( j \) for a pair of n.n. spins has the general form

\begin{equation}
\label{43}
P_{j}(\overrightarrow{S}_{i}+\overrightarrow{S}_{i+1})=\prod _{l\neq j}\frac{\left[ l(l+1)-\overrightarrow{S}^{2}\right] }{\left[ l(l+1)-j(j+1)\right] }
\end{equation}
 where \( \overrightarrow{S}=\overrightarrow{S}_{i}+\overrightarrow{S}_{i+1} \)
. For the AKLT model, \( j=2 \) and \( l=1,0 \) . From (42) and (43),

\begin{equation}
\label{44}
H_{AKLT}=\sum _{i}\left[ \frac{1}{2}(\overrightarrow{S}_{i}.\overrightarrow{S}_{i+1})+\frac{1}{6}(\overrightarrow{S}_{i}.\overrightarrow{S}_{i+1})^{2}+\frac{1}{3}\right] 
\end{equation}
The method of construction of the AKLT Hamiltonian can be extended to higher
spins and to dimensions d > 1. The MG Hamiltonian (apart from a numerical prefactor
and a constant term) can be written as 

\begin{equation}
\label{45}
H=\sum _{i}P_{\frac{3}{2}}(\overrightarrow{S}_{i}+\overrightarrow{S}_{i+1}+\overrightarrow{S}_{i+2})
\end{equation}
The \( S=1 \) HAFM and the AKLT chains are in the same Haldane phase, characterised
by a gap in the excitation spectrum. The physical picture provided by the VBS
ground state of the AKLT Hamiltonian holds true for real systems \cite{46}.
The excitation spectrum of \( H_{AKLT} \) cannot be determined exactly. Arovas
et al. \cite{47} have proposed a trial wave function

\begin{equation}
\label{46}
\left| k\right\rangle =N^{-\frac{1}{2}}\sum ^{N}_{j=1}e^{ikj}S^{\mu }_{j}\left| \psi _{VBS}\right\rangle ,\mu =z,+,-
\end{equation}
 and obtained 

\begin{equation}
\label{47}
\epsilon (k)=\frac{\left\langle k\left| H_{VBS}\right| k\right\rangle }{\left\langle k\mid k\right\rangle }=\frac{25+15cos(k)}{27}
\end{equation}
The gap in the excitation spectrum \( \Delta =\frac{10}{27} \) at k = \( \pi  \).
Another equivalent way of creating excitations is to replace a singlet by a
triplet spin configuration \cite{48}.

\section{Spin Ladders}

A. Undoped ladders

In the last Section, we discussed some exact results for interacting spin systems.
The powerful technique of BA was described. The BA cannot provide knowledge
of correlation functions. There is another powerful technique for 1d many body
systems known as bosonization \cite{49} which enables one to calculate various
correlation functions for 1d systems. After the discovery of high-\( T_{c} \)
cuprate systems, the study of 2d AFMs acquired considerable importance. There
are, however, not many rigorous results available for 2d spin systems. Ladder
systems interpolate between a single chain (1d) and the square lattice (2d)
and are ideally suited for the study of the crossover from 1d to 2d. Consider
a two-chain spin ladder (Figure 4) described by the AFM Heisenberg exchange
interaction Hamiltonian 
\begin{equation}
\label{48}
H_{J-J_{R}}=\sum _{\left\langle ij\right\rangle }J_{ij}\overrightarrow{S}_{i}.\overrightarrow{S}_{j}
\end{equation}
The n.n. intra-chain and the rung exchange interactions are of strength \( J \)
and \( J_{R} \) respectively. When \( J_{R} \) = 0, one obtains two decoupled
AFM spin chains for which the excitation spectrum is known to be gapless. Dagotto
et al. \cite{50} derived the interesting result that the lowest excitation
spectrum is separated by an energy gap from the ground state. The result is
easy to understand in the simple limit in which the exchange coupling \( J_{R} \)
along the rungs is much stronger than the exchange coupling \( J \) along the
chains. The intra-chain coupling may thus be treated as perturbation. When \( J=0 \)
, the exact ground state consists of singlets along the rungs, each singlet
having the spin configuration \( \frac{1}{\sqrt{2}}\left[ \left| \uparrow \downarrow \right\rangle -\left| \downarrow \uparrow \right\rangle \right]  \).
The ground state energy is \( -\frac{3J_{R}N}{4} \), where \( N \) is the
number of rungs in the ladder. In first order perturbation theory, the correction
to the ground state energy is zero. The ground state has total spin \( S=0 \).
A \( S=1 \) excitation may be created by promoting one of the rung singlets
to a \( S=1 \) triplet. A triplet has the spin configuration \( \left| \uparrow \uparrow \right\rangle  \)
(\( S^{z}=+1 \)), \( \frac{1}{\sqrt{2}}\left| \uparrow \downarrow +\downarrow \uparrow \right\rangle  \)
(\( S^{z}=0 \) ) and \( \left| \downarrow \downarrow \right\rangle  \) (\( S^{z}=-1 \)).
A triplet costs an exchange energy equal to \( J_{R} \) . The weak coupling
along the chains gives rise to a band of propagating \( S=1 \) magnons with
the dispersion relation 

\begin{equation}
\label{49}
\omega (k)=J_{R}+Jcosk
\end{equation}
in first order perturbation theory (\( k \) is the wave vector). The spin gap,
defined as the minimum excitation energy is given by

\begin{equation}
\label{50}
\Delta _{SG}=\omega (\pi )\simeq (J_{R}-J)
\end{equation}
The two-spin correlations decay exponentially along the chains showing that
the ground state is a quantum spin liquid (QSL). As the rung exchange coupling
\( J_{R} \) decreases, one expects that the spin gap will also decrease and
ultimately become zero at a critical value of \( J_{R} \). Barnes et al. \cite{51},
however, put forward the conjecture that \( \Delta _{SG}>0 \) for all \( \frac{J_{R}}{J} \)
> 0, including the isotropic limit \( J_{R}=J \). A variety of numerical techniques
like exact diagonalization of finite-sized ladders \cite{50}, Quantum Monte
Carlo (QMC) simulations \cite{52} and density-matrix renormalization group
(DMRG) \cite{53} have verified the conjecture. We now consider the case of
an n-chain ladder. A surprising fact emerging out of several theoretical studies
\cite{54,55} is: the excitation spectrum has spin gap (is gapless) when n is
even (odd). In the first (second) case, the two-spin correlation function has
an exponential (power-law) decay. For odd n, the ladder has properties similar
to those of a single chain. The strong coupling limit (\( J_{R}\gg J \)) again
provides a physical picture as to why this is true. When n is even, the S =
\( \frac{1}{2} \) spins along a rung continue to form a singlet ground state.
Hence the creation of a S =1 excitation requires a finite amount of energy as
in the case of the two-chain ladder. The gap should decrease as n increases
so that the gapless square lattice limit is reached for large n. When n is odd,
each rung consists of an odd number of spins, each of magnitude \( \frac{1}{2} \)
. The inter-rung (intra-chain) coupling \( J \) generates an effective interaction
between the \( S=\frac{1}{2} \) rung states, which because of rotational invariance,
should be of the Heisenberg form with an effective coupling \( J_{eff} \) setting
the energy scale. The equivalence of an odd-chain ladder to the single Heisenberg
chain leads to a gapless excitation spectrum. Rojo \cite{56} has given a rigorous
proof of the gaplessness of the excitation spectrum when n is odd. Khveshchenko
\cite{57} has shown that for odd-chain ladders, a topological term governing
the dynamics at long wavelengths appears in the effective action, whereas, it
exactly cancels for even-chain ladders. The topological term has similarity
to the one that causes the difference between integer and half-odd integer spin
chains. In the first case, the spin excitation spectrum has the well-known Haldane
gap. In the latter case, the LSM theorem shows that the excitation spectrum
is gapless. Ghosh and Bose \cite{58} have constructed an n-chain spin ladder
model for which the exact ground state can be determined for all values of n.
For n even (odd), the excitation spectrum has a gap (is gapless). This is true
even for large n, thus the square lattice limit cannot be reached in the model.
Thermodynamic properties of the \( S=\frac{1}{2} \) two-chain ladder have been
first studied by Troyer et al. \cite{59}. Using a quantum transfer matrix method,
they obtained reliable results down to temperature \( T\sim 0.2J \). The AFM
correlation length \( \xi _{AFM} \) has been found to be 3-4 lattice spacings.
The magnetic susceptibility \( \chi (T) \) shows a crossover from a Curie-weiss
form, \( \chi (T)=\frac{C}{T+\theta } \)at high temperature to an exponential
fall-off, \( \chi (T)\sim \frac{e^{-\frac{\Delta _{SG}}{T}}}{\sqrt{T}} \) as
\( T\rightarrow 0 \). The fall-off is a signature of a finite spin gap \( \Delta _{SG} \).
Frischmuth et al. \cite{60}, using a powerful loop algorithm, have calculated
the magnetic susceptibility and found evidence for the gapped (gapless) excitation
spectrum in the case of an even (odd)-chain ladder.

A major interest in the study of ladder systems arises from the fact that there
is a large number of experimental realizations of ladder systems. A comprehensive
review of major experimental systems is that by Dagotto \cite{61}. We discuss
here only a few interesting ladder systems. Hiroi et al. \cite{62} were the
first to synthesize the family of layer compounds \( Sr_{n-1}Cu_{n+1}O_{2n} \).
Rice et al. \cite{54} subsequently recognized that these compounds contained
weakly-coupled ladders of \( \frac{n+1}{2} \) chains. For n =3 and 5, respectively,
one gets the two-chain and three-chain ladder compounds. Azuma et al. \cite{63}
have determined the temperature dependence of the magnetic susceptibility in
these ladder compounds experimentally. A spin gap is indicated by the sharp
fall of \( \chi (T) \)for \( T<300K \) in the two-chain ladder compound \( SrCu_{2}O_{3} \)
. The magnitude of the spin gap is \( \Delta _{SG}\sim 420K \). This is approximately
in agreement with the theoretical result of \( \Delta _{SG}\simeq \frac{J}{2} \),
if an exchange coupling \( J\sim 1200K \) is assumed. For the three-chain ladder
compound \( Sr_{2}Cu_{3}O_{5} \) , Azuma et al. found that \( \chi (T) \)
approaches a constant as \( T\rightarrow 0 \), , as expected for the 1d Heisenberg
AFM chain. Muon spin relaxation measurements by Kojima et al. \cite{64} shows
the existence of a long range ordered state with N\'{e}el temperature \( T_{N} \)
= 52 K, brought about by the interlayer coupling. No sign of long range ordering
was found in the two-chain ladder compound, confirming the difference between
odd and even chain ladders. The compound \( LaCuO_{2.5} \) is formed by an
array of weakly interacting two-chain ladders \cite{65}. The evidence of spin-liquid
formation at intermediate temperatures (confirmed by the existence of a spin
gap) and an ordered N\'{e}el state at low temperatures, shows that the spin
singlet state is in close competition with a N\'{e}el state. Spin ladders, belonging
to the organic family of materials, have also been synthesized. A recent example
is the compound \( (C_{5}H_{12}N)_{2}CuBr_{4} \) \cite{66}. This compound
is a good example of a strongly coupled (\( \frac{J_{R}}{J}\simeq 3.5 \) )
ladder system. The phase diagram of the AFM spin ladder in the presence of an
external magnetic field is particularly interesting. In the absence of the magnetic
field and at \( T=0 \), the ground state is a quantum spin liquid with a gap
in the excitation spectrum. At a field \( H_{c_{1}} \), there is a transition
to a gapless Luttinger liquid phase (\( g\mu _{B}H_{c_{1}}=\Delta _{SG} \)
, the spin gap, \( \mu _{B} \) is the Bohr magneton and g the Land\'{e} splitting
factor). There is another transition at an upper critical field \( H_{c_{2}} \)
to a fully polarised FM state. Both \( H_{c_{1}} \) and \( H_{c_{2}} \) are
quantum critical points. The phase transitions that occur at these points are
quantum phase transitions as they occur at \( T=0 \). At a quantum critical
point, the system switches from one ground state to another. The transition
is brought about by changing a parameter (magnetic field in the present example)
other than temperature. At small temperatures, the behaviour of the system is
determined by the crossover between two types of critical behaviour: quantum
critical behaviour at \( T=0 \) and classical critical behaviour at \( T\neq 0 \).
Quantum effects are persistent in the crossover region at small finite temperature
and such effects can be probed experimentally. Refs. \cite{67,68,69} give extensive
reviews of quantum critical phenomena. In the case of the ladder system \( (C_{5}H_{12}N)_{2}CuBr_{4} \),
the magnetization data, obtained experimentally, exhibit universal scaling behaviour
in the vicinity of the critical fields, \( H_{c_{1}} \) and \( H_{c_{2}} \)
. We remember that in the vicinity of critical points, physical quantities of
a system exhibit scaling behaviour. Quantum spin systems provide several examples
of quantum phase transitions and organic spin ladders are systems which provide
experimental testing grounds of theories of such transitions. For inorganic
spin ladder systems, the value of \( H_{c_{1}} \) is too high to be experimentally
accessible.\\
B. Frustrated spin ladders

Bose and Gayen \cite{70} have studied a two-chain spin ladder model with frustrated
diagonal couplings (Figure 5, frustrated spin systems are defined in Section
4). The intra-chain and diagonal spin-spin interactions are of equal strength
\( J \). The exchange interactions along the rungs are of strength \( J_{R} \)
. It is easy to show that for \( J_{R}\geq 2J \), the exact ground state consists
of singlets along the rungs with the energy \( E_{g}=-\frac{3J_{R}N}{4} \)where
\( N \) is the number of rungs. Xian \cite{71} pointed out that the Hamiltonian
of the frustrated ladder model can be written as

\begin{equation}
\label{51}
H=J_{R}\sum _{i}\overrightarrow{S}_{1i}.\overrightarrow{S}_{2i}+J\sum _{i}\overrightarrow{P}_{i}.\overrightarrow{P}_{i+1}
\end{equation}
 where, \( \overrightarrow{P_{i}}=\overrightarrow{S}_{1i}+\overrightarrow{S}_{2i} \)
, represents a composite operator at the i-th rung and `1' and `2' refer to
the lower and upper chains respectively. Due to the commutativity of the rung
interaction part of the Hamiltonian with the second part, the eigenstates of
the Hamiltonian can be described in terms of the total spins of individual rungs.
The energy eigenvalue for the state with singlets on all the rungs is \( E^{s}_{g}=-\frac{3J_{R}N}{4} \).
The second term in the Hamiltonian (Eq.(51)) does not contribute in this case.
If the two rung spins form a triplet, the second term is equivalent to the Hamiltonian
of a spin-1 Heisenberg chain with a one-to-one correspondence between a rung
of the ladder and a site of the \( S=1 \) chain. Because the two parts of H
commute, the eigenvalue, when the rung spins form a triplet is

\begin{equation}
\label{52}
E^{T}_{g}=(Je_{0}+\frac{J_{R}}{4})N
\end{equation}
 where \( e_{0}=-1.40148403897(4) \) is the ground state energy/site of the
spin 1 Heisenberg chain. Comparing the energy \( E^{T}_{g} \) with the energy
\( E^{s}_{g} \) of the rung singlet state, one finds that as long as \( \frac{J_{R}}{J}>(\frac{J_{R}}{J})_{c}=e_{0} \)
, the latter state is the exact ground state. At the critical value \( (\frac{J_{R}}{J})_{c} \)
, there is a first order transition from the rung singlet state to the Haldane
phase of the \( S=1 \) chain. The lowest spin excitation in the rung singlet
state can be created by replacing a rung singlet (\( S=0 \)) by a triplet (\( S=1 \)).
The triplet excitation spectrum has no dynamics. In a more general parameter
regime, i.e., when the intra-chain exchange interaction is not equal in strength
to the diagonal exchange interaction, the ground and the excited states can
no longer be determined exactly. In this case, one takes recourse to approximate
analytical and numerical methods. Kolezhuk and Mikeska \cite{72} have constructed
a class of generalised \( S=\frac{1}{2} \) two-chain ladder models for which
the ground state can be determined exactly. The Hamiltonian \( H \) is a sum
over plaquette Hamiltonians and each plaquette Hamiltonian contains various
two-spin as well as four-spin interaction terms. They have further introduced
a toy model, the Generalised Bose-Gayen (GBG) model which has a rich phase diagram
in which the phase boundaries can be determined exactly. Recently, some integrable
spin ladder models with tunable interaction parameters have been introduced
\cite{73,74,75}. The integrable models, in general, contain multi-spin interaction
terms besides two-spin terms.\\
C. Doped spin ladders

A major reason for the strong research interest in ladders is that doped ladder
models are toy models of strongly correlated systems. The most well-known examples
of the latter are the high-\( T_{c} \) cuprate systems. As already mentioned
in the Introduction, these systems exhibit a rich phase diagram as a function
of the dopant concentration. Doping effectively replaces the spin-\( \frac{1}{2} \)
's associated with the \( Cu^{2+} \) ions in the \( CuO_{2} \) planes by holes.
The holes are mobile in a background of antiferromagnetically interacting Cu
spins. Also, due to strong Coulomb correlations, the double occupancy of a site
by two electrons, one with spin up and the other with spin down, is prohibited.
This is a non-trivial many body problem because it involves a competition between
two processes: hole delocalization and exchange energy minimization. A hole
moving in an antiferromagnetically ordered spin background, say, the N\'{e}el
state, gives rise to parallel spin pairs which raise the exchange interaction
energy of the system. The questions of interest are: whether a coherent motion
of the holes is possible, whether two holes can form a bound state (in the superconducting
(SC) phase of the doped cuprates, charge transport occurs through the motion
of bound pairs of holes), the development of SC correlations, the possibility
of phase separation of holes etc. For the cuprates, a full understanding of
many of these issues is as yet lacking (see \cite{76} for a recent review of
high-\( T_{c} \) superconductivity). The doped ladders are simple model systems
in which the consequences of strong correlation can be studied with greater
rigour than in the case of the structurally more complex cuprate systems. Recent
experimental evidence \cite{61} suggests that some phenomena are common to
ladder and cuprate systems. The study of ladder systems is expected to provide
insight on the common origin of these phenomena. Some ladder compounds can be
doped with holes. Much excitement was created in 1996 when the ladder compound
\( Sr_{14-x}Ca_{x}Cu_{24}O_{41} \) was found to become SC under pressure at
x = 13.6. The transition temperature \( T_{c}\sim 12K \) at a pressure of 3
GPa. As in the case of SC cuprate systems, holes form bound pairs in the SC
phase of ladder systems. The possibility of binding of hole pairs in a two-chain
ladder system was first pointed out by Dagotto et al. \cite{50}. The strongly
correlated doped ladder system is described by the t-J Hamiltonian

\begin{equation}
\label{53}
H_{t-J}=-\sum _{\left\langle ij\right\rangle ,\sigma }t_{ij}(\widetilde{C}^{+}_{i\sigma }\widetilde{C}_{j\sigma }+H.C.)+\sum _{\left\langle ij\right\rangle }J_{ij}(\overrightarrow{S}_{i}.\overrightarrow{S}_{j}-\frac{1}{4}n_{i}n_{j})
\end{equation}
 The \( \widetilde{C}^{+}_{i\sigma } \)and \( \widetilde{C}_{i\sigma } \)
are the electron creation and annihilation operators which act in the reduced
Hilbert space (no double occupancy of sites).

\begin{eqnarray}
\widetilde{C}^{+}_{i\sigma } & = & C^{+}_{i\sigma }(1-n_{i-\sigma })\nonumber \\
\widetilde{C_{i\sigma }} & = & C_{i\sigma }(1-n_{i-\sigma })\label{54} 
\end{eqnarray}
 \( \sigma  \) is the spin index and \( n_{i} \) , \( n_{j} \) are the occupation
numbers of the i-th and j-th sites respectively. The term proportional to \( n_{i}n_{j} \)
is often dropped. The first term in Eq.(53) describes the motion of holes with
hopping integrals \( t_{R} \) and \( t \) for motion along the rung and chain
respectively. In the conventional \( t-J \) ladder model, i and j are n.n.
sites. The second term (minus the \( -\frac{1}{4}n_{i}n_{j} \) term) is the
usual AFM Heisenberg exchange interaction Hamiltonian. The \( t-J \) model
thus describes the motion of holes in a background of antiferromagnetically
interacting spins. In the undoped limit, each site of the ladder is occupied
by a spin-\( \frac{1}{2} \) and the \( t-J \) Hamiltonian reduces to the AFM
Heisenberg Hamiltonian. Removal of a spin creates an empty site, i.e., a hole.
A large number of studies have been carried out on \( t-J \) ladder models.
These are reviewed in Refs. \cite{61,77,78}. We describe briefly some of the
major results. A hole-doped single AFM chain is an example of a Luttinger Liquid
(LL) which is different from a Fermi liquid. The latter describes interacting
electron systems in higher dimensions and at low temperatures. A novel characteristic
of a LL is spin-charge separation due to which the charge and spin parts of
an electron (or hole) move with different velocities and thus become separated
in space. The undoped two-chain ladder has a spin gap. This gap remains finite
but changes discontinuously on doping. This is because there are now two disctinct
triplet excitations (remember that the spin gap is the difference in energies
of the lowest triplet excitation and the ground state). One triplet excitation
is obtained by exciting a rung singlet to a rung triplet as in the undoped case.
A new type of triplet excitation is obtained in the presence of two holes. A
clear physical picture is obtained in the limit \( J_{R}\gg J \). In this case,
the ground state predominantly consists of singlets along the rungs. On the
introduction of a hole, a singlet spin pair is broken and the hole exists with
a free spin-\( \frac{1}{2} \) . In the presence of two holes on two separate
rungs, the two free spin-\( \frac{1}{2} \) 's combine to give rise to an excited
triplet state. The ground state is a singlet and consists of a bound pair of
holes. The binding of holes can be understood in a simple manner. Two holes
located on two different rungs break two rung singlets and the exchange interaction
energy associated with two rungs is lost in the process. If the holes are located
on the same rung, the exchange interaction energy of only one rung is lost.
If \( J_{R} \) is much greater than the other parameters of the system, the
holes preferentially occupy rungs in pairs. As \( J_{R} \) decreases in strength,
the hole bound pair has a greater spatial extent. The lightly doped ladder system
is not in the LL phase, i.e., no spin-charge separation occurs. The system is
in the so-called Luther-Emery phase with gapless charge excitations and gapped
spin excitations. A variety of numerical studies show that the hole pairs and
the spin gap are present even in the isotropic limit \( J_{R}=J \) . Also,
the relative state of hole pairs has approximate ``d-wave'' symmetry with
the pairing amplitude having opposite signs along the rungs and the chains.
The d-wave symmetry is a feature of strong correlation and is considered to
be the symmetry of the pairing state in the case of cuprate systems. 

Bose and Gayen \cite{70,79,80} have constructed a two-chain \( t-J \) ladder
model with frustrated diagonal couplings. The intra-chain n.n. and the diagonal
hopping integrals have the same strength \( t \). The other parameters have
been defined earlier. The special structure of the model enables one to determine
the exact ground and excited states in the cases of one and two holes. The most
significant result is an exact, analytic solution of the eigenvalue problem
associated with two holes in the infinite \( t-J \) ladder. The binding of
holes has been explicitly demonstrated and the existence of the Luther-Emery
phase established. For conventional \( t-J \) ladders (the diagonal bonds are
missing), the only exact results that have been obtained are through numerical
diagonalization of finite-sized ladders. Derivation of exact, analytical results
in this case has not been possible so far. The reason for this is that as a
hole moves in the antiferromagnetically interacting spin background, spin excitations
in the form of parallel spin pairs are generated. Proliferation of states with
spin excitations makes the solution of the eigenvalue problem extremely difficult.
In the case of the frustrated \( t-J \) ladder model, there is an exact cancellation
of the terms containing parallel spin pairs \cite{80}. Thus the hole has a
perfect coherent motion through the spin background. Frahm and Kundu \cite{81}
has constructed an integrable \( t-J \) ladder model and obtained the phase
diagram. The model contains terms describing correlated hole hopping in chains
which may not be realizable in real systems. Several studies have been carried
out on the two-chain Hubbard ladder as well as on multi-chain Hubbard and \( t-J \)
ladders. References of some of the studies may be obtained from \cite{61}.

\section{Frustrated spin models in 2d}

In Sections 2 and 3 we have discussed quasi-1d interacting spin systems, namely,
spin chains and ladders. As already mentioned in the Introduction, the \( CuO_{2} \)
plane of the undoped cuprate systems is a 2d AFM. The undoped cuprates exhibit
AFM LRO below a N\'{e}el temperature \( T_{N} \) . On the introduction of a
few percent of holes, the AFM LRO is rapidly destroyed leaving behind spin-disordered
states in the \( CuO_{2} \) planes. This fact has triggered lots of interest
in the study of spin systems with spin disordered states as ground states. Frustrated
spin models are ideal candidates for such systems. To understand the origin
of frustration, consider the AFM Ising model on the triangular lattice. An elementary
plaquette of the lattice is a triangle. The Ising spin variables have two possible
values, \( \pm 1 \), corresponding to up and down spin orientations. An antiparallel
spin pair has the lowest interaction energy \( -J \) . A parallel spin pair
has the energy \( +J \) . In an elementary triangular plaquette, there are
three interacting spin pairs. Due to the topology of the plaquette, all the
three pairs cannot be simultaneously antiparallel. There is bound to be at least
one parallel spin pair. The parallel spin pair may be located along any one
of the three bonds in the plaquette and so the ground state is triply degenerate.
The Ising model on the full triangular lattice has a highly degenerate ground
state such that the entropy/spin is a finite quantity. As a result, the system
never orders including at \( T=0 \) . Frustration occurs in the system since
all the spin pair interaction energies cannot be simultaneously minimised. On
the other hand, consider the AFM Ising model on the square lattice. All the
four spin pairs in an elementary square plaquette can be made antiparallel and
so there is no frustration. The system exhibits magnetic order below a critical
temperature. If one of the spin pair interactions in each elementary square
plaquette is FM and the rest AFM, frustration occurs in the square lattice spin
system. A spin system with mixed FM and AFM interactions is frustrated if the
sign of the product of exchange interactions around an elementary plaquette
is negative. In the case of a purely AFM model, frustration occurs if the number
of bonds in an elementary plaquette of the lattice is odd. Examples of such
lattices in 2d are the triangular and kagom\'{e} lattices. In 3d, the pyrochlore
lattice, the elementary plaquette of which is a tetrahedron provides an example.
A spin system is also frustrated due to the presence of both n.n. as well as
further-neighbour interactions. Consider AFM n.n. as well as n.n.n. interactions
between a row of three Ising spins. Again, all the three spin pairs cannot simultaneously
be made antiparallel.

Let us now treat the spins as classical vectors (\( S\rightarrow \infty  \)).
For AFM spin-spin interaction, the lowest energy is achieved for an antiparallel
spin configuration. In the classical limit, the spins on a bipartite lattice
are ordered in the AFM N\'{e}el state. On a non-bipartite lattice, such as the
triangular lattice, the classical ground state represents a compromise between
competing requirements. In the ground state, the spins form an ordered three-sublattice
structure with \( 120^{0} \) between n.n. spins on different sublattices. The
ground state of the classical Heisenberg model on the kagom\'{e} lattice is,
however, highly degenerate and disordered. We now consider the full quantum
mechanical spin Hamiltonian and ask the question how the classical ground states
are modified when quantum fluctuations are taken into account. In the case of
the triangular lattice, it is now believed that the quantum mechanical ground
state of the \( S=\frac{1}{2} \) HAFM model has AFM LRO of the N\'{e}el-type,
i.e., quantum fluctuations do not destroy the three-sublattice order of the
classical ground state. In the second scenario, when the classical ground state
is highly degenerate and disordered, thermal/quantum fluctuations select a subset
of states which tend to incorporate some degree of long range order. This is
the phenomenon of `order from disorder' which is counterintuitive since order
is brought about by fluctuations which normally have disordering effects. The
classical kagom\'{e} lattice HAFM ground states include both coplanar as well
as noncoplanar spin arrangements and fluctuations lead to the selection of coplanar
order. This kind of ordering is particularly true for large values of the spin
S. As the magnitude of the spin is decreased towards \( S=\frac{1}{2} \), the
quantum fluctuations increase in strength. These fluctuations often destroy
the ordered structure obtained for large S. The quantum mechanical ground states
of the \( S=\frac{1}{2} \) HAFM on the kagom\'{e} and pyrochlore lattices have
been found to be spin disordered. Some recent references of frustrated magnetic
systems are \cite{82,83}. The triangular lattice \( S=\frac{1}{2} \) HAFM
is the first example of a spin model in which frustration occurs due to lattice
topology \cite{84}. The \( S=\frac{1}{2} \) HAFM model has also been studied
on a partially frustrated pentagonal lattice \cite{85} and a parameter region
identified in which the ground state has AFM LRO of the N\'{e}el-type.

Two well-known examples of spin-disordered states are the quantum spin liquid
(QSL) and dimer or valence bond (VB) states. A QSL state is a spin singlet with
total spin \( S=0 \) and has both spin rotational and translational symmetry.
In a VB state, spin rotational symmetry is present but tanslational symmetry
is broken. In such states pairs of spins form singlets which are called VBs
or dimers with the VBs being frozen in space. A well-known example of a QSL
state is the resonating-valence-bond (RVB) state \cite{84} which is a coherent
linear superposition of VB states (Figure 6). The RVB state is the starting
point for the well-known RVB theory of high-\( T_{c} \) SC. Spin-disordered
(no AFM LRO as defined in Eq. (8)) states with novel order parameters are:\\
(a) Chiral states \\
In these states, the spins are arranged in configurations characterised by the
order parameter

\begin{equation}
\label{55}
\Delta _{i}=\left\langle \overrightarrow{S}_{i}.(\overrightarrow{S}_{i+\widehat{x}}\times \overrightarrow{S}_{i+\widehat{y}}\right\rangle 
\end{equation}
 with the three spins belonging to one plaquette of the square lattice and \( \widehat{x},\widehat{y} \)
denoting unit vectors in the x and y directions respectively. The chiral state
breaks time reversal symmetry or a reflection about an axis (parity).\\
(b) Dimer states\\
These are the VB states in which the VBs are frozen in space. A well-known example
of such states is the columnar dimer (CD) states. In such states, the VBs are
arranged in columns. On the square lattice, four such states are possible. The
order parameter of CD states is

\begin{equation}
\label{56}
D_{l}=\left\langle \eta (l)\overrightarrow{S}_{l}.(\overrightarrow{S}_{l+\widehat{x}}+i\overrightarrow{S}_{l+\widehat{y}}-\overrightarrow{S}_{l-\widehat{x}}-i\overrightarrow{S}_{l-\widehat{y}})\right\rangle 
\end{equation}
where the l-sites are even and \( \eta (l)=+1(-1) \) if both \( l_{x} \) and
\( l_{y} \) are even (odd). The order parameter takes the values \( 1,i,-1,-i \)
for the four CD states shown in Figure 3. \\
(c) Twisted states:\\
At the classical level (\( S\rightarrow \infty  \)), the spins in the twisted
state are arranged in incommensurate structures. These configurations can be
visualised as spins lying in a plane and with a twist angle in some direction.
It is possible that such states survive the inclusion of quantum fluctuations.
The order parameter is vectorial in nature and is given by 

\begin{equation}
\label{57}
T_{l}=\left\langle \overrightarrow{S}_{l}\times (\overrightarrow{S}_{l+\widehat{x}}+\overrightarrow{S}_{l+\widehat{y}})\right\rangle 
\end{equation}
These states are called spin nematics and are different from helimagnets in
which both \( T_{l} \) and the spin-spin correlation functions show LRO.\\
(d) Strip or Collinear States\\
In a classical picture, the spins are ferromagnetically ordered in the \( x \)
direction and antiferromagnetically ordered in the y direction. The configuration
obtained by rotating the previous one by \( \frac{\pi }{2} \) is also possible.
The order parameter is given by

\begin{equation}
\label{58}
C_{l}=\left\langle \overrightarrow{S}_{l}.(\overrightarrow{S}_{l+\widehat{x}}-\overrightarrow{S}_{l+\widehat{y}}+\overrightarrow{S}_{l-\widehat{x}}-\overrightarrow{S}_{l-\widehat{y}})\right\rangle 
\end{equation}
\( C_{l} \) takes the values \( 1,-1 \) in the two different strip configurations.

The spin-disordered states decribed above are quantum-coherent states and are
characterised by novel order parameters. The term `quantum paramagnet' is often
used to describe such states.

Examples of real frustrated systems are many \cite{82,83}. The best studied
experimental kagom\'{e} system is the magnetoplumbite, \( SrCr_{8-x}Ga_{4+x}O_{19} \)
. The system consists of dense kagom\'{e} layers of \( S=\frac{3}{2} \) Cr
ions, separated by dilute triangular layers of Cr. In a mean-field theory of
the HAFM, the high temperature susceptibility is given by 

\begin{equation}
\label{59}
\chi =\frac{C}{T+\theta _{cw}},T\gg T_{N}
\end{equation}
 Here, the Curie constant \( C=\frac{\mu ^{2}_{B}p^{2}}{3k_{B}} \) , where
\( \mu _{B} \) is the Bohr magneton and, \( p=g[s(s+1)]^{\frac{1}{2}} \),
\( g \) being the Land\'{e} splitting factor governing the splitting of the
spin multiplet in a magnetic field. Also, \( \theta _{cw} \) is the Curie-Weiss
temperature. The N\'{e}el ordering temperature \( T_{N} \) is defined experimentally
using bulk probes. One looks for singularities in either the specific heat C(T)
or the temperature derivative of the susceptibility \( \chi (T) \) . In the
case of non-frustrated systems, \( T_{N}\sim \theta _{cw} \) and in the second
case, \( T_{N}<<\theta _{cw} \) . Since a frustrated system may not order at
all in a conventional sense, the hallmark of such a system is \( T_{c}<<\theta _{cw} \)
where \( T_{c} \) is the temperature below which new types of spin order set
in. The Curie-weiss temperature \( \theta _{cw} \) is an experimentally measurable
quantity. One defines an empirical measure of frustration by the quantity

\begin{equation}
\label{60}
f=-\frac{\theta _{cw}}{T_{c}}
\end{equation}
 Frustration corresponds to \( f>1 \) . For the kagom\'{e} AFM \( SrCr_{8}Ga_{4}O_{19} \)
, f is as high as 150. 

\( SrCrGaO \) displays unconventional low-T behaviour \cite{86}. One of these
is the insensitivity of the specific heat \( C(T) \) to applied magnetic fields
\( H \) as large as twice the temperature. Susceptibility measurements show
the existence of a gap \( \Delta _{SG} \) in the triplet spin excitation spectrum.
For \( T<\Delta _{SG},\chi \sim e^{-\frac{\Delta _{SG}}{k_{B}T}} \). The specific
heat, however, does not decrease exponentially for \( T<\Delta _{SG} \), i.e.,
does not have a thermally activated behaviour. It has a \( T^{2} \) dependence.
This fact along with the experimental observation of insensitivity of \( C(T) \)
to external magnetic field have been explained by suggesting that a large number
of singlet excitations fall within the triplet gap \cite{86} . Numerical evidence
of such excitations has been obtained in the case of the \( S=\frac{1}{2} \)
HAFM on the kagom\'{e} lattice \cite{87}. The number of such excitations has
been found to be \( \sim (1.15)^{N} \) where N is the number of spins in the
lattice. Mambrini and Mila \cite{88} have recently established that a subset
of short-range RVB states captures the specific low energy physics of the kagom\'{e}
lattice HAFM and the number of singlet states in the singlet-triplet gap is
\( (1.15)^{N} \) in agreement with the numerical results. The appearance of
singlet states in the singlet-triplet gap could be a generic feature of strongly
frustrated magnets. Other examples of such systems are: the \( S=\frac{1}{2} \)
frustrated HAFM on the \( \frac{1}{5} \)-depleted square lattice describing
the 2d AFM compound \( CaV_{4}O_{9} \) \cite{89,90}, the HAFM on the 3d pyrochlore
lattice and a 1d system of coupled tetrahedra \cite{91}. Bose and Ghosh \cite{90}
have constructed a frustrated \( S=\frac{1}{2} \) AFM model on the \( \frac{1}{5}- \)depleted
square lattice and have shown that in different parameter regimes the plaquette
RVB (PRVB) and the dimer states are the exact ground states. In the PRVB state,
the four-spin plaquettes (Figure 7) are in a RVB spin configuration which is
a linear superposition of two VB states. In one such state, the VBs (spin singlets)
are horizontal and in the other state the VBs are vertical. In the dimer state,
VBs or dimers form along the bonds joining the four-spin plaquettes. Both the
PRVB and the dimer states are spin disordered states. The state intermediate
between the PRVB and the dimer states has AFM LRO. Both the PRVB and dimer phases
are characterised by spin gaps in the excitation spectrum. For the unfrustrated
HAFM model on the \( \frac{1}{5} \) -depleted square lattice, Troyer et al.
\cite{92} have carried out a detailed study of the quantum phase transition
from an ordered to a disordered phase. In the ordered phase, the excitation
spectrum is gapless. The spin gap \( \Delta _{SG} \) continuously goes to zero
in a power-law fashion at the quantum critical point separating a gapped disordered
phase from a gapless ordered phase. Chung et al. \cite{93} have carried out
an extensive study on the possible paramagnetic phases of the Shastry-Sutherland
model \cite{19}. In addition to the usual dimer phase, they find the existence
of a phase with plaquette order and also a topologically ordered phase with
deconfined \( S=\frac{1}{2} \) spinons and helical spin correlations. Takushima
et al. \cite{94} have studied the ground state phase diagram of a frustrated
\( S=\frac{1}{2} \) quantum spin model on the square lattice. This model includes
both the Shastry-Sutherland model as well as the spin model on the \( \frac{1}{5}- \)depleted
square lattice as special cases. The nature of quantum phase transitions among
the various spin gap phases and the magnetically ordered phases has been clarified. 

Lieb and Schupp \cite{95} have derived some exact results for the fully frustrated
HAFM on a pyrochlore checkerboard lattice in 2d. This lattice is a 2d version
of the 3d pyrochlore lattice. The ground states have been rigorously proved
to be singlets. The Lieb-Mattis theorem is applicable only for bipartite lattices.
Hence, the proof for the pyrochlore checkerboard lattice is a new result. Lieb
and Schupp have further proved that the magnetization in zero external field
vanishes separately for each frustrated tetrahedral unit. Also, the upper bound
on the susceptibility is \( \frac{1}{8} \) in natural units for both \( T=0 \)
and \( T\neq 0 \) . Frustration can also occur from a competition between exchange
anisotropy and the transverse field terms as in the case of the transverse Ising
model. Moessner et al. \cite{96} have shown that the transverse Ising model
on the triangular (kagom\'{e}) lattice has an ordered (disordered) ground state. 

Frustrated AFMs with short-range dimer or RVB states as ground states have a
gap in the spin excitation spectrum and the two-spin correlation function has
an exponential decay as a function of the distance separating the spins. A single
branch describes the \( S=1 \) triplet excitation spectrum. In Section 2 we
pointed out that in the case of the \( S=\frac{1}{2} \) HAFM chain, a pair
of spinons (each spinon has spin \( S=\frac{1}{2} \) ) are the fundamental
excitations. The lowest excitation spectrum is thus not a single branch of \( S=1 \)
magnon excitations but a continuum of scattering states with well-defined lower
and upper boundaries. AFM compounds in 2d, in general, have excitation spectra
described by \( S=1 \) magnons. Anderson \cite{84} suggested that a RVB state
may support pairs of spinons as excitations which are deconfined via a rearrangement
of the VBs. In this case, an extended and highly dispersive continuum of excitations
is expected. Recently, Coldea et al. \cite{97} have investigated the ground
state ordering and dynamics of the 2d \( S=\frac{1}{2} \) frustrated AFM \( Cs_{2}CuCl_{4} \)
using neutron scattering in high magnetic fields. The dynamic correlations exhibit
a highly dispersive continuum of excited states which are characteristic of
the RVB state and arise from pairs of \( S=\frac{1}{2} \) spinons. A recent
paper `RVB Revisited' by P.W.Anderson \cite{98} focuses on the relevance of
the RVB state to describe the normal state of the \( CuO_{2} \) planes in the
high-\( T_{c} \) cuprate superconductors.

\section{Concluding remarks}

In Sections 1-4, a brief overview of low-dimensional quantum magnets, specially,
antiferromagnets has been given. The subject of quantum magnetism has witnessed
an unprecedented growth in research activity in the last decade. This is one
of the few research areas in which rigorous theories can be worked out and experimental
realizations are not diffficult to find. Coordination chemists and material
scientists have prepared novel materials and constructed new application devices.
Experimentalists have employed experimental probes of all kinds to refine old
data and uncover new phenomena. Theorists have taken recourse to a variety of
analytical and numerical techniques to explain the experimentally observed properties
as well as to make new predictions. Powerful theorems have been proposed and
exact results obtained. This trend is still continuing and will lead to further
breakthroughs in materials, phenomena and techniques in the coming years. The
study of doped magnetic materials which include cuprates, ladders and spin chains
has acquired considerable importance in recent times. The dopants can be magnetic
and nonmagnetic impurities as well as holes. To give one example, it has been
possible to dope the spin-1 Haldane-gap AFM compound \( Y_{2}BaNiO_{5} \) with
holes. Neutron scattering experiments reveal the existence of midgap states
and an incommensurate double-peaked structure factor \cite{99}. Several new
experimental results on the Haldane-gap AFMs have been obtained in the last
few years \cite{100} which add to the richness of phenomena observed in magnetic
syatems. In this overview, only a few of the aspects of quantum magnetism have
been highlighted. Conventional 2d and 3d magnets have not been discussed at
all as a good understanding of these materials already exists. We have not discussed
recent advances in material applications which include materials exhibiting
colossal magnetoresistance, molecular magnets, nanocrystalline magnetic materials,
magnetoelectronic devices (the study of which constitutes the new subject of
spintronics) etc. A report on some of these developments as well as some recent
issues in quantum magnetism may be obtained from Ref. \cite{101}.

\newpage 

Figure Captions

Figure 1. The Shastry-Sutherland model

Figure 2. Five types of interaction in the \( J_{1}-J_{2}-J_{3}-J_{4}-J_{5} \)
model

Figure 3. Four columnar dimer states

Figure 4. A two-chain spin ladder

Figure 5. The two-chain frustrated spin ladder model

Figure 6. An example of a RVB state 

Figure 7. The \( \frac{1}{5}- \)depleted square lattice. 

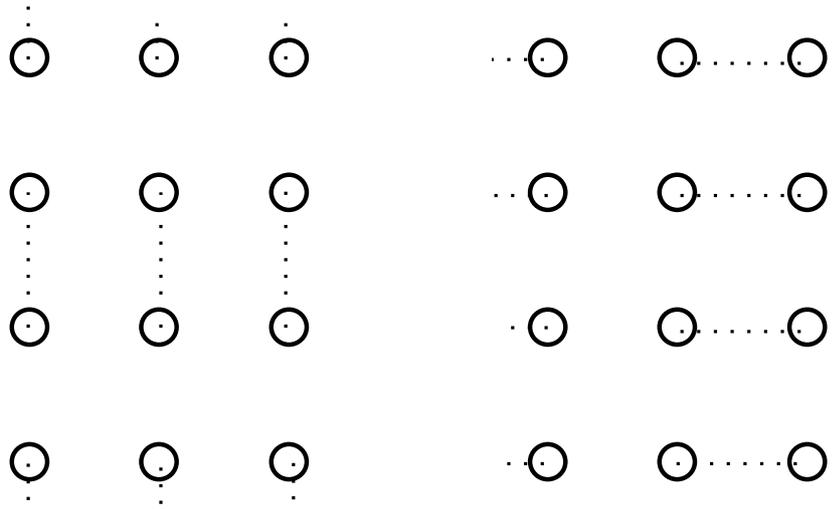
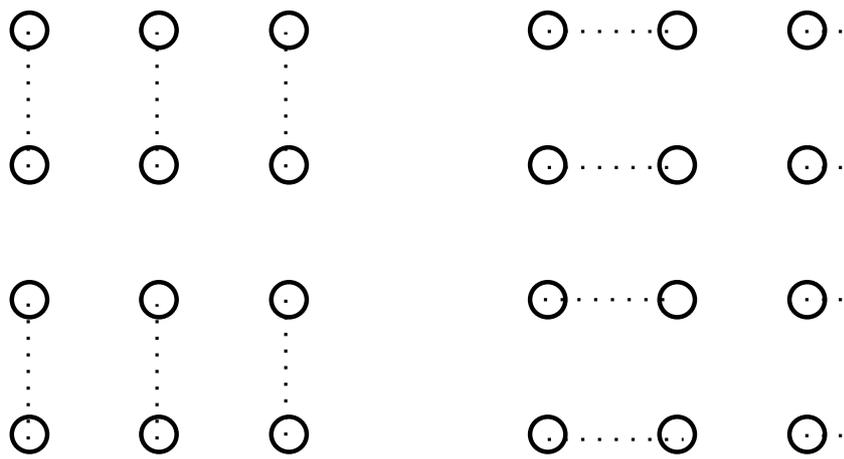
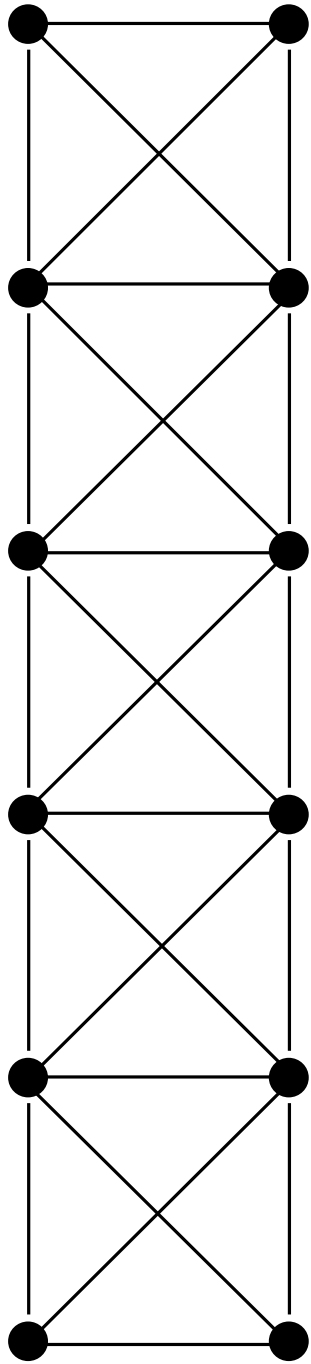
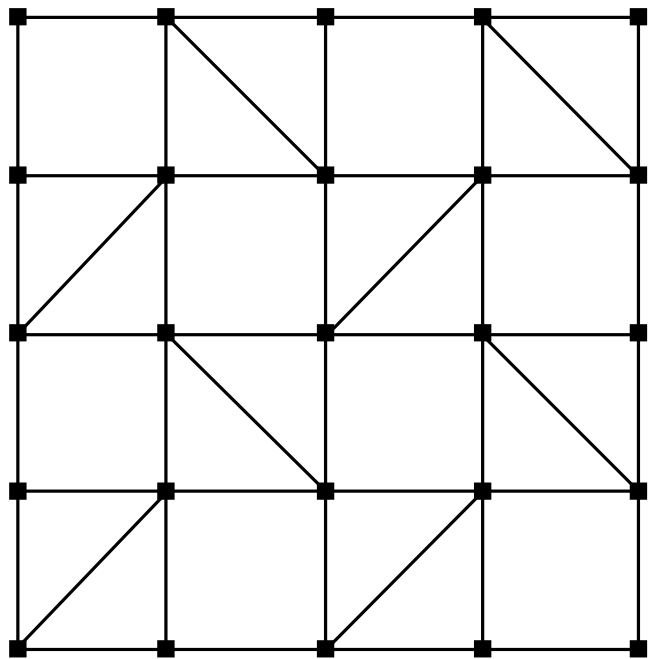
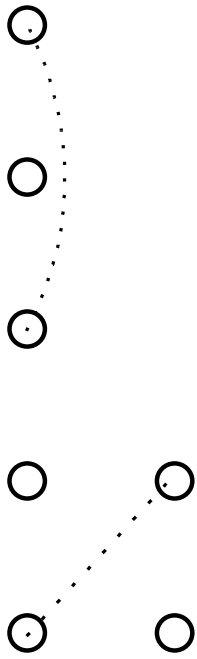
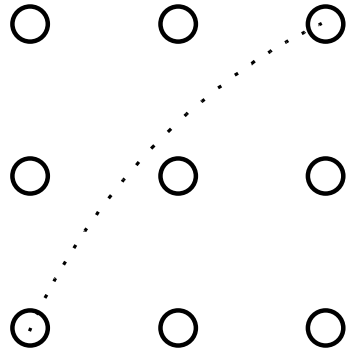
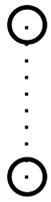
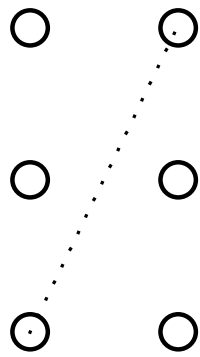
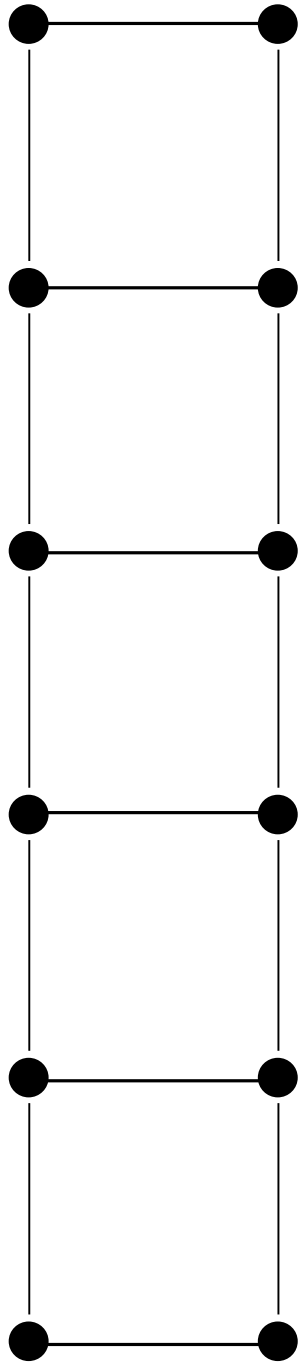
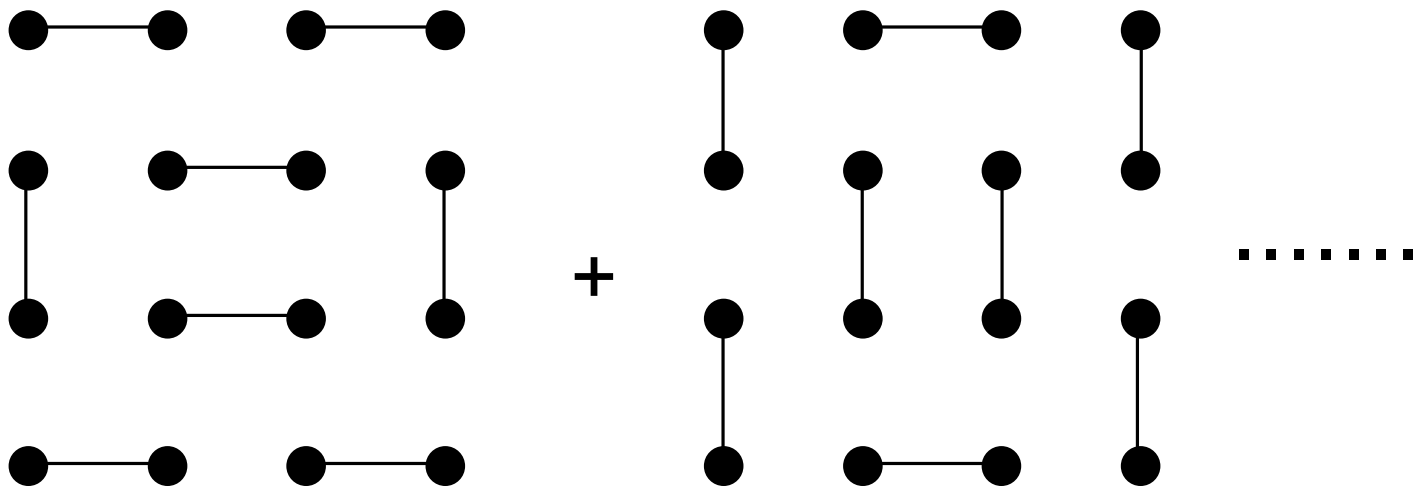
\end{document}